\documentclass[%
 reprint,
 amsmath,amssymb,
 aps,
]{revtex4-2}

\usepackage{graphicx}
\usepackage{dcolumn}
\usepackage{bm}

\begin{document}

\newcommand{\scr}[1]{_{\mbox{\protect\scriptsize #1}} }

\preprint{APS/123-QED}

\title{Mode Converting Alfv\'{e}n Waves from Magnetic Reconnection Enhancing the Energy Source for the  Aurora Borealis}

\author{Harsha Gurram}
\email{hgurram@wisc.edu}
\author{Jan Egedal}%
\affiliation{%
 University of Wisconsin-Madison, Wisconsin, USA
}%
\author{William Daughton}
\affiliation{%
 Los Alamos National Laboratory, USA
}%
\date{\today}
\begin{abstract}
Previous studies have concluded that the Hall magnetic field structures generated during magnetic reconnection are carried away by kinetic Alfv\'{e}n waves (KAW). Here we apply a kinetic simulation with an ion/electron mass ratio closer to its natural value and find that much-reduced damping rates permit the KAW to convert into shear Alfv\'{e}n waves (SAW). For magnetotail reconnection these SAW provides efficient transport of wave energy, enhancing the energy input for the Aurora Borealis by orders of magnitude above previous estimates.   
\end{abstract}
\maketitle

Magnetic reconnection plays a fundamental role in nearly all magnetized plasmas as it allows the magnetic field lines to change topology in collisionless plasmas, and it thereby controls the spatial and temporal evolution of explosive phenomena such as solar flares, coronal mass ejections and magnetic storms in Earth's magnetotail. It further enables magnetic energy to be converted into high speed flows, thermal energy and waves -- for example energy released from magnetic reconnection sites in the Earth's magnetotail have been postulated as a possible source of energy for the Aurora Borealis. Although the energy of the ion exhaust jets are locally significant  they do not transport the energy at sufficiently large distances to reach Earth \cite{Birn2005EnergyRA,Birn2009}.  Instead, wave energy propagating along magnetic field lines is potentially  a more important mediator in the transfer of energy released in the magnetotail to the auroral zones \cite{keiling_02}.

It is well known that during magnetic reconnection a quadrupolar out-of-plane or Hall magnetic field is formed by the field aligned currents that flow in the vicinity of the magnetic separatrices when ion and electrons decouple on the length scales less than the ion inertial length ($d_i$) \cite{sonnerup:1979}. These Hall fields and their associated magnetic flux are swept from the reconnection layers super-Alfv\'{e}nically by kinetic Alfv\'{e}n waves (KAW) \cite{Shay_2011,Rogers_2001}. Observations from Cluster spacecraft have confirmed that such waves are indeed generated by magnetic reconnection and play a vital role in transporting energy \cite{chaston_05,chaston_09, Gershman_2017}.  
Recent work, however, brings into question the role of KAWs in powering the aurora. 
The studies in Refs.~\cite{Sharma_2018,Shay_2011} suggest that the KAWs, sought to carry the Hall field at large distances, become nearly fully attenuated over magnetotail distances of $\sim20R_e$ (with a loss of about $99\%$ of the wave energy). Furthermore, polar spacecraft observations at $4 -6 R_e$ radial distance \cite{Wygant_02,keiling_02,Keiling_2003}, and Cluster observations at $\sim 5 R_E$ radial distance \cite{Vaivads_2004} suggest that it is not energy associated with the KAW that powers the aurora. Rather, shear Alfv\'{e}n waves (SAW) are found to be adequate to account for the auroral brightening at locations magnetically conjugate to the satellites. As a possible resolutions to these issues, in this letter, using a high-fidelity numerical simulation,  we demonstrate how reconnection generated KAWs are converted into SAWs providing near unattenuated transfer of a significant fraction of the magnetic energy released in the magnetotail.

\begin{figure}[!ht]
\includegraphics[width=0.9\linewidth]{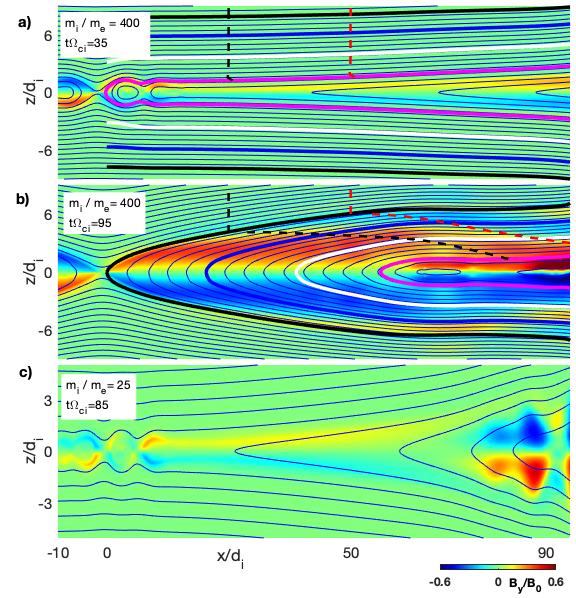}
\caption{\label{fig:figby} Hall structure at times a) $35\Omega_{ci}^{-1}$ b) $95\Omega_{ci}^{-1}$ and c) $85 \Omega_{ci}^{-1}$ in the simulation with $m_i/m_e = 400$, $400$ and $25$. In (a) and (b) magenta, white, blue and black solid lines represents fieldline reconnecting at $t \Omega_{ci}=35, 55, 75 $ and $95$,  black and red dashed lines represents cuts at $<x> = 25d_i$ and $40d_i$.}
\end{figure}

 Applying a fully kinetic simulation we investigate how the Hall magnetic field structure generic to reconnection propagates away at large distance from the reconnection site. For the implementation of a large domain, numerical tractability can be accomplished by introducing a reduced ion to electron mass ratio (for 2D domains the numerical cost scales as $(m_i/m_e)^{5/2}$). For example, results in Ref.~\cite{Shay_2011} was done for a small domain ($\sim 50d_i$) at $m_i/m_e$ ranging from 25 to 400, and large domain ($ \geq200d_i$) simulation results in Ref.~\cite{Sharma_2018} were obtained at $m_i/m_e = 25$. It was demonstrated that the Hall magnetic field structures propagates as KAWs, however, to encapsulate the evolution of the Hall fields a large domain simulation with large mass ratio becomes necessary. This is because a value of $m_i/m_e=25 $ renders the simulation KAW over-damped through Landau damping on the electrons. In this paper we present results from a simulation implemented at $m_i/m_e=400$. Similar to magnetospheric conditions, the thermal speed of the electrons is then much larger than the Alfv\'{e}n speed, rendering Landau damping negligible. This dramatically influences the quantitative and qualitative structure, hence fate of the KAWs as they carry the Hall magnetic field away from the X-line region, permitting mode-conversion to SAWs. 

Our fully kinetic simulation of reconnection utilizes the VPIC code \cite{bowers:2009} and is initialized with a  Harris equilibrium without an out of plane guide field. We have a simulation domain of 200\textit{d$_i$} X 30\textit{d$_i$} = 20480 cells X 3072 cells and open boundary conditions \cite{daughton:2007} that allows the exhaust to fully develop without any spurious boundary effects. Considering typical magnetotail parameters of $B=20 nT$, $n=0.1 cm^{-3}$, $T_i =1 keV$, $T_e =300 eV$ this domain translates to $25R_e$ X  $4R_e$ and we study the nature of the wave propagation as far as $90 d_i\sim10R_e$ from the X-point.The 2D simulation has an initial unperturbed magnetic field given by \textit{B$_y$}=\textit{B$_0\tanh$(z/d$_i$)}, based on the peak Harris density \textit{n$_0$} with the $B_0=0.5$. The results presented have the following parameters: \textit{m$_i$/m$_e$}=400, initial uniform temperature satisfying $T_{i0}/T_{e0}$=5, $\omega_{pe}/\omega_{ce}$=2, and a background density $n_b=0.5 n_0$. With these parameters the upstream total plasma $\beta$= 0.3. To make contact with previous results we also include a lower mass ratio run performed at $m_i/m_e=25$ with all the other parameters unchanged.

 Figs.~\ref{fig:figby}(a,b)  show the Hall field $B_y$ profile at times $t\Omega_{ci}=35$ and $t\Omega_{ci}=95$ during the reconnection process. The Hall structure extends hundreds of $d_i$ and is significant throughout the majority of the region between the separatrices and center of the exhaust, filling an exhaust region much larger than any length scales typically associated with Hall physics. The profiles in Fig.~\ref{fig:figby}(c) is for a similar time as that in Fig.~\ref{fig:figby}(b) and although a similar amount of magnetic flux has been reconnected, this run at $m_i/m_e=25$ does not develop the prominent $B_y$ structures seen in the $m_i/m_e=400$ run. The magenta, white, blue and the black lines in Fig.~\ref{fig:figby}(a) represent the fieldlines reconnecting at $t\Omega_{ci}=35$, $55$, $75$ and $95$ respectively hence marking contours of constant $\psi$'s. As the reconnection progresses the fieldlines travel into the exhaust, where their length is reduced due to breaking and reconnecting as shown in Fig.~\ref{fig:figby}(b). To examine the waves setting the structure of the reconnection region we study the profiles of Hall field along particular fieldlines as these propagate through the simulation domain. As such, the Hall field $B_y$ is evaluated as a function of $(x,t)$ for constant values of $\psi$, where $\psi$ is the flux function defined as $\textbf{B}=\hat{\textbf{y}}\times\nabla{\psi}+B_y\hat{\textbf{y}}$.
 
 \begin{figure*}[!ht]
\includegraphics[width=0.54\linewidth]{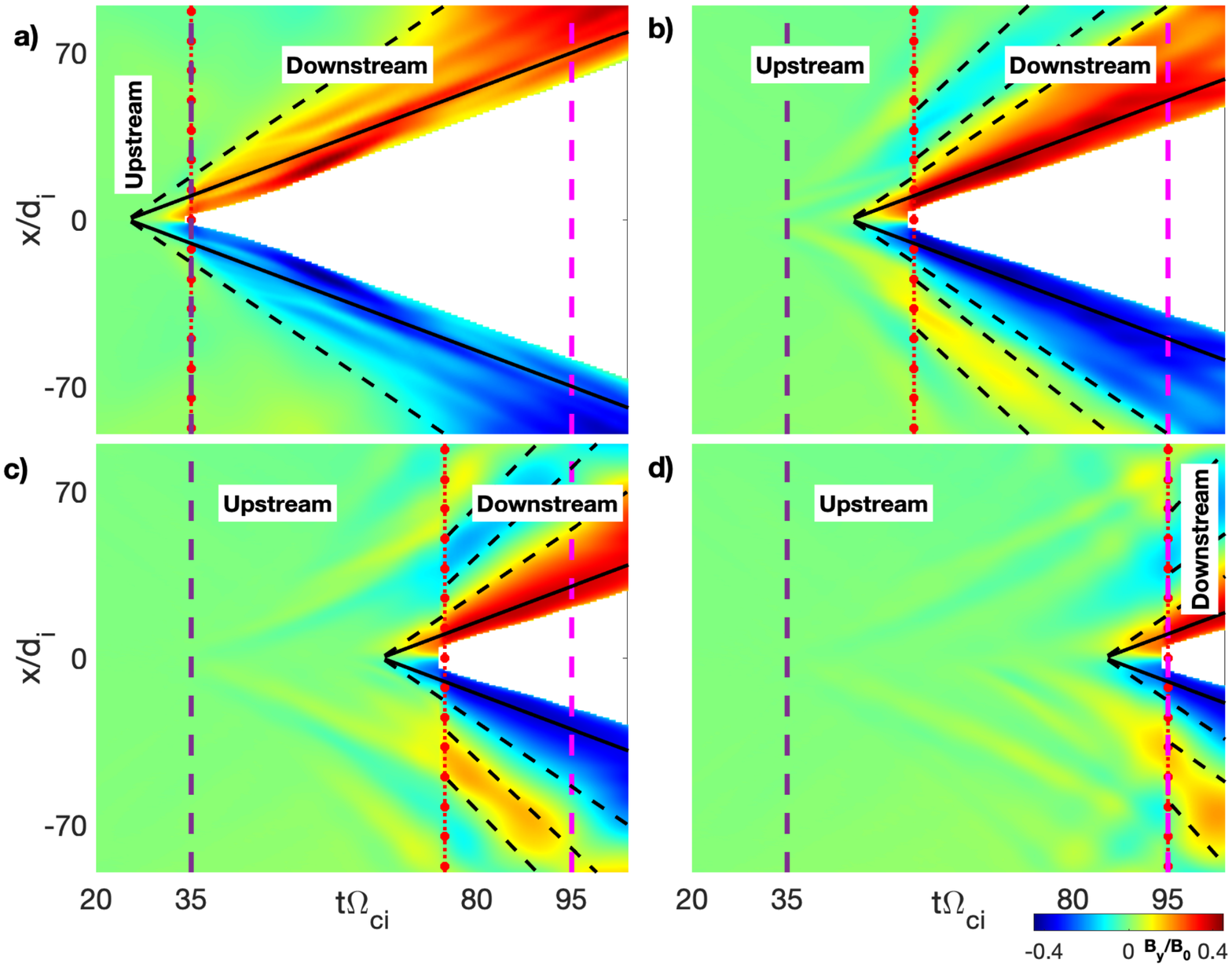}
\includegraphics[width=0.34\linewidth]{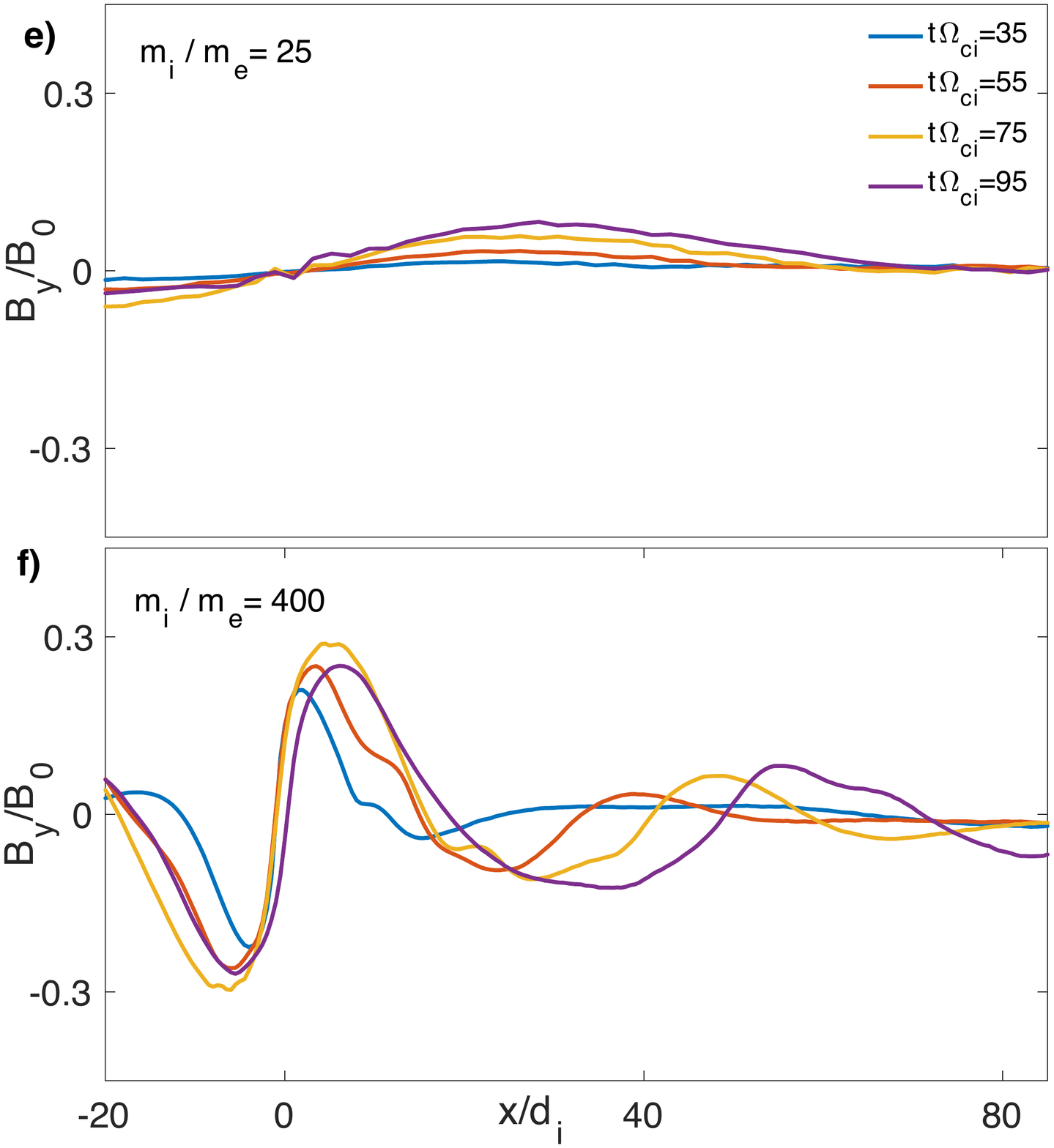}
\caption{\label{figbyl} Time evolution of B$_y$ field along the fieldline that reconnects at a) $t\Omega_{ci}=35$ b) $t\Omega_{ci}=55$ c) $t\Omega_{ci}=75$ and d) $t\Omega_{ci}=95$ represented by magenta, white, blue and black fieldline respectively in Fig.~\ref{fig:figby}, with red dashed line representing the separatrix. From top to center slope m is for (a)1.8 and 1, (b) \& (c) 2.5, 1.8 and 1 (d) 2.75, 2.5, 1.8 and 1. e) and f) The Hall field evolution along the separatrix at various times for the two runs at
$m_i/m_e$ = 25 and $m_i/m_e$ = 400, respectively. The differences between e) and f) show that the KAW at  $m_i/m_e$ = 25 is over-damped.}
\end{figure*}

  To obtain the profiles of $\left. B_y(x,t)\right|_{\psi}$ we trace the Hall field, $B_y$ along the moving field lines (characterized by fixed values of $\psi$) before, at the time it reconnects, and after, \textit{i.e.} in the inflow, the separatrix, and in the exhaust  from time $t =20\Omega^{-1}_{ci}$ to $104\Omega^{-1}_{ci}$. Repeating this for the other $\psi$-values selected in Fig.~\ref{fig:figby}, we generate several profiles of $ \left. B_y(x,t)\right|_{\psi}$ as shown in Fig.~\ref{figbyl}(a-d).  As fieldlines travel in the exhaust we observe a $x_{min}$, corresponding to the value of $x$ for $z=0$ at the selected value of $\psi$; the white spaces in the profiles represent nonphysical values $|x|<x_{min}$. In Fig.~\ref{fig:figby}(a) the magenta fieldline $\psi_1$, has $x_{min}=0$ at $t\Omega_{ci}=35$ and $x_{min}=20d_i$ at  $t\Omega_{ci}=75$, illustrating how  $x_{min}$ moves with the outflow at a speed $\sim V_{a}$. 

 The vertical red dotted dashed line in Fig.~\ref{figbyl}(a-d) represents the separatrix at that time-step. The left and right of the separatrix correspond to the inflow and the exhaust regions, respectively.  The peak $B_y$ structure developed at the separatrix in Fig.~\ref{figbyl}(a) spreads along the fieldlines in the exhaust as seen in Fig.~\ref{figbyl}(b-d). Additionally, multiple $B_y$ structures develop in the exhaust as reconnection progresses as evident in Fig.~\ref{figbyl}(e) which shows the $B_y$ at the separatrix at various times. These new structures generated at the separatrix, propagate at speed greater than the peak $B_y$ in the exhaust. This is apparent from the slope of the peak $B_y$ structure in $ \left. B_y(x,t)\right|_{\psi}$ which is $\sim 1$ whereas the new structures are steeper and have slope ranging from $2.5-3$ times the peak $B_y$ structure. The Hall fields are hence carried by waves which originate at the separatrix and propagate both parallel and transverse to the fieldlines at various speeds. We first focus our study on the propagation of the peak structures along the fieldline and determine the wave speed.

The primary wavenumber $k_{\parallel}d_{i}$ associated with the parallel $B_y$ propagation is determined by Fourier transform at the separatrix,  where $\parallel$ is $w.r.t$ to the magnetic fieldline, yielding $k_{\parallel}d_{i}= 0.3$ at $t\Omega_{ci}=35$ and $0.13$ at $t\Omega_{ci}=95$. As the reconnection develops we see a reduction in $k_{\parallel}d_i$ as the formation of the multiple features along the fieldlines enhances the parallel wavelengths.  An estimate for the velocity of these waves can be determined from the slope of  $B_y$ structure shown by the dashed lines in Figs.~\ref{figbyl}(a-d). Here the slope of the peak $B_y$  is $1.8$ giving us $V_{sim}=dl/dt=1.8d_i\Omega_{ci}=1.8V_{a}$ as estimate for the speed of the peak $B_y$. In previous studies at low  $m_i/m_e$  \cite{Sharma_2018,Shay_2011} only a single peak in $B_y$ was observed (as reproduced by our simulation in Fig.~\ref{figbyl}(e) at $m_i/m_e=25$). In contrast, in our simulation for $m_i/m_e=400$ in Fig.~\ref{figbyl}(f) contains  Hall fields  with additional wave structures along the fieldline and a $\sim300\%$ increase in the peak amplitude. Thus, in runs with low $m_i/m_e$ the waves carrying the Hall fields are over-damped due to electron Landau damping, whereas in our run at $m_i/m_e=400$ the Landau damping is small  and  changes qualitatively the wave dynamics in the exhaust over large distance. 
 \begin{figure}[!ht]
\includegraphics[width=0.9\linewidth]{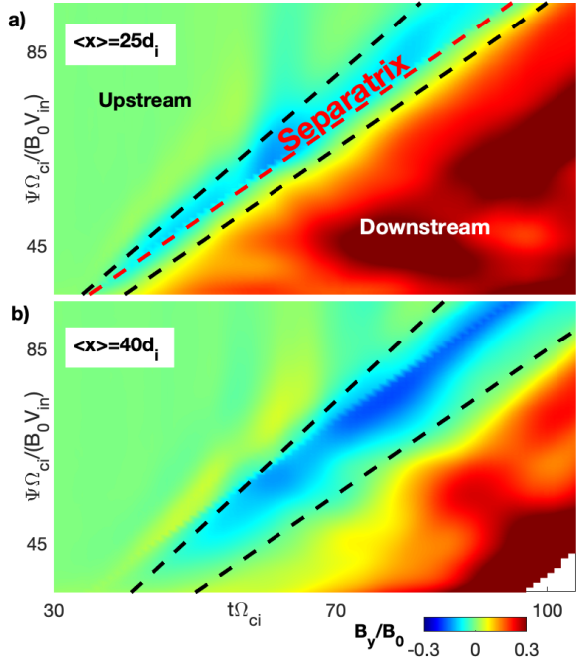}
\caption{\label{figbyperp} Profile of the Hall field evolution perpendicular to the fieldlines (a)$B_y(\psi,t,\left<x\right>=25d_i)$ and (b)$B_y(\psi,t,\left<x\right >=40d_i)$. The slope of the lines characterizing the structure in (a) and (b) 1.4, 1.0.}
\end{figure}

In Figs.~\ref{figbyl}(c,d) faint $B_y$ structures are observed to the left of the separatrix which are  formed by the waves that travel transverse to the fieldlines. To estimate $k_{\bot}$, we examine $B_y$ as a function of $(\psi,t)$ by taking a cut  perpendicular to the fieldlines at a fixed distance along the field line, $\left < x \right > = x- x_{min}$. We do this for two different values of $\left < x \right >=25d_i$ and $40d_i$, which will aid our study of waves near and far away from X-point. These cuts are represented in Fig.~\ref{fig:figby}(a,b) by the black and red dashed line for $25d_i$ and $40d_i$ respectively.

 Fig.~\ref{figbyperp} shows the evolution of the Hall structure $B_y(\psi,t,\left < x \right >$=25$d_i)$ and $B_y(\psi,t,\left < x \right >$=$40d_i)$. %
 A fieldline reconnecting at  $t\Omega_{ci}=a$ has $\psi=a\,B_0V_{in}/\Omega_{ci}$ (taking $\psi=0$ at $t=0$). For $\left < x \right > =25 d_i$, in 
Fig.~\ref{figbyperp} (a) we see a main Hall field of $\sim -0.1$  for most of the time after the fieldlines reconnect (the separatrix is shown by red dash-dotted line). We also see a positive $B_y$ structure forming at $t\Omega_{ci}=35$ on the fieldline that reconnects at that time step, this structure is seen by field lines at larger values of $\psi$ before they reconnect. This is evidence that waves carry these structure transversely into the inflow. The broadening of the structure shows  that the transverse propagation velocity is greater than the inflow speed  $V_{in}$ of the plasma. Similar structures are also seen at $\left < x \right >=40d_i$, but the structures have reduced amplitude compared to those at smaller $\left < x \right >$. As mentioned above,  the fieldlines get shortened after reconnection, explaining the  the white patch in the lower right corner of  Fig.~\ref{figbyperp}(b).

The perpendicular phase velocity of the waves are estimated from the slope of the lines in Fig.~\ref{figbyperp}(a) relative  to the slope of the red dash dotted line of the separatrix. While the slope of the separatrix line is $1B_0V_{in}/\Omega_{ci}$, the slope of the main structure  in ($\psi$,t) is $\sim 1.4$, such that $V_{\perp}\approx 1.4V_{in}=0.14V_{a0}$. Taking a horizontal cut at any $\psi$ and performing Fourier transform in time yield $\omega= 0.31\Omega_{ci}$ at small $\left < x \right >$ and $\omega=0.153 \Omega_{ci}$ at large $\left < x \right >$, and hence we have the $k_{\perp}d_{i}=\omega/V_{\perp}=2.2$ and $1.09$, respectively.

Using a standard two-fluid plasma model, in the low frequency limit 
we analyze the branch of waves associated with Alfv\'{e}n waves and kinetic Alfv\'{e}n waves. 
\begin{equation}
\frac{\omega^2}{k_{\parallel}^2}=V_a^2[1+(\frac{\beta}{1+\beta})k^2d_i^2]\approx V_a^2[1+k^2\rho_s^2]  
\end{equation}
 where $\rho_s^2=\beta d_i^2$. The above dispersion relation corresponds to the Kinetic Alfv\'{e}n Wave for $k_{\perp}^2\rho_s^2\gg 1$ and the Shear Alfv\'{e}n Wave for $k_{\perp}^2\rho_s^2\ll 1$. The waves carrying the Hall fields are oblique as they have $k_{\perp}\gg k_{\parallel}$, hence $k^2d_i \approx k_{\perp}^2d_{i}$. For the waves originating close to the separatrices we use the plasma parameters from the simulation $(B=0.7; n=0.8; T_i+T_e=0.5$, $\beta =0.3)$, and find  $kd_{i}=1.37$ and $2.7$. From the values above for  $kd_{i}$ we get $k_{\perp}^2\rho_s^2= 0.55$ for   $\left<x\right>=25d_i$ and  $2.2$ for $\left<x\right>=40d_i$. The predicted value of the velocity of these waves from Eq.~1  is then $\omega/k_{\parallel}=1.8V_a$ close to the X-point and $1.2V_a$ as we move away from the X-point into the exhaust. The Hall structure of reconnection is thus carried by waves which are oblique with $k_{\perp} \gg k_{\parallel}$. As the waves propagate into the exhaust their perpendicular lengths increases, and consequently $k_\perp d_i$ decreases in the exhaust leading to an overall decrease in the parallel wave speed. The wave speeds are super Alfv\'{e}nic near the X-point but become on the order of the Alfv\'{e}n speed deeper into the exhaust ($\sim 60d_i$ from the X-point).

Whenever $k_\perp\rho_i> 1$, the wave supports fluctuations in electric field which enable transfer of energy between the wave fields and plasma particles via Landau resonance \cite{Barnes_1966, Hasegawa_1976, Stix}. Landau damping is not important when the perpendicular wavelength is larger than the ion acoustic gyroradius and the electron inertial length ($k_\perp\rho_s$ and $k_\perp d_e < 1$ ). For $ \beta > m_e/m_i$, the Landau damping becomes a sensitive function of the perpendicular wavelength and waves with $k_\perp\rho_s \gg 1$ are strongly  damped \cite{Lysak_1996}. The rapid mode conversion from KAW to SAW is therefore important as it eliminates the energy damping of 99\%, expected for the KAWs propagating towards Earth \cite{Sharma_2018}. 

\begin{figure}[!ht]
\includegraphics[width=\linewidth]{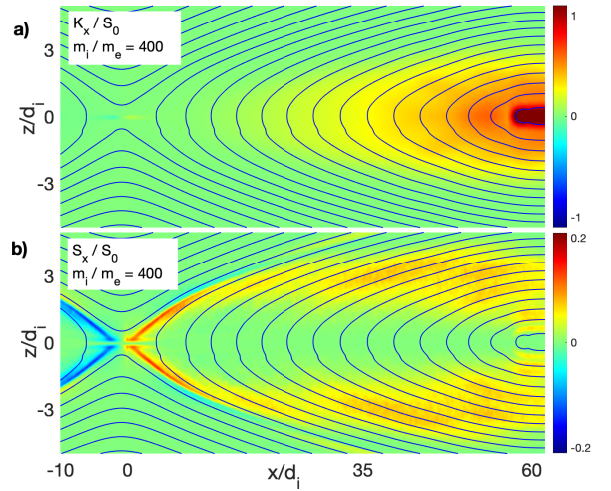}	
\caption { At time $t\Omega_{ci}=90$ parallel a) bulk kinetic energy flux $K_x  ( = 0.5\rho v^2 v_x)$ b) Poynting flux $S_x$ associated with the waves wherein both fluxes are normalized by  $S_0 = cB_{0}^{2}/\mu_0$. Note that $S_x$ is calculated in the frame of the ion outflow with a speed of $V\scr{exhaust}= \pm V_a$.}
\label{figsx}
\end{figure}

To determine the wave energy and energy transport we calculate Poynting Flux in the frame of reference of the ion exhaust by performing a Lorentz transformation. Doing so we exclude contributions from any other fluxes embedded in the outflow, giving us the relevant component of the Poynting flux vector as:  $S\cdot\hat{b}_{xz}\approx S_x=(E\times B)_x/\mu_0\approx E_zB_y/\mu_0$ (where E's and B's are in the frame of motion of the exhaust). Fig.~\ref{figsx}(b) shows the calculated Poynting flux of wave-fields at time $t\Omega_{ci}=90$. For magnetotail parameters, the energy associated with this wave travels distances of  $\gg 60d_i\,\,(9R_e)$ without significant attenuation. The large amplitudes of these waves, up to $60\%$ of the reconnecting field, carry an amount of energy corresponding to about $20\%$ of the bulk kinetic energy in (shown in Fig~\ref{figsx}(a)) away from the x-line. In contrast, for the previous scenario of  KAWs without mode-conversion the  Poynting flux drops drastically (by 99\%) before reaching $\sim 5R_e$. Furthermore, at $m_i/m_e=400$ the initial amplitude of the KAWs is a factor of 3 above the amplitude at $m_i/m_e=25$. Thus, the Poynting flux transmitted to the ionosphere by mode conversion to SAWs is enhanced by a factor of about $10^3$ above previous estimates. Based on the previous estimate it was argued that  magnetotail reconnection can directly drive the Auroral phenomenon \cite{Sharma_2018}, and with the above  significant increase in the transmitted power this only becomes more likely. Additionally, this conversion becomes relevant in solar corona where SAWs propagate unattenuated, carrying energy to distances comparable to the flare loop length. 

In summary, the quadrupole Hall structure of reconnection, which stretch the magnetic field in the out-of-plane direction, is first carried away from the reconnection site by kinetic Alfv\'{e}n waves. Due to their dispersive nature, as the KAW propagates their wavenumber, $k$,  decreases, corresponding to a mode conversion into shear Alfv\'{e}n waves. The Hall fields hence propagate along the magnetic field super-Alfv\'{e}nically close to the X-line and becomes Alfv\'{e}nic at large distances. The wave dynamics are associated with a substantial Poynting flux away from the X-line, enhanced by a factor of $10^3$ above estimates in previous work. This may lead to particle acceleration \cite{Wygant_02} and help account for auroral brightening at locations magnetically conjugate to spacecraft observations \cite{Watt_10}. 



\nocite{*}
\bibliography{apssamp}

\end{document}